\documentclass[superscriptaddress,aps,preprintnumbers,amsmath,amssymb,nofootinbib,10pt,twocolumn,prd, floatfix]{revtex4-2}
\usepackage{graphicx,subfigure, xcolor}
\usepackage[bookmarksopen,colorlinks=true,linkcolor=blue,citecolor=blue,urlcolor=blue]{hyperref}
\usepackage{ulem}

\begin{document}
\preprint{RIKEN-iTHEMS-Report-24, YITP-24-10, CTPU-PTC-24-04}

\title{Resurgence in Lorentzian quantum cosmology: No-boundary saddles and \\ resummation of quantum gravity corrections around tunneling saddle points}

\author{Masazumi Honda}
\affiliation{Interdisciplinary Theoretical and Mathematical Sciences Program (iTHEMS), RIKEN, Wako 351-0198, Japan,\\
Theoretical Sciences Visiting Program, Okinawa Institute of Science and Technology Graduate University (OIST), Onna, 904-0495, Japan
}

\author{Hiroki Matsui}
\affiliation{Center for Gravitational Physics and Quantum Information, 
Yukawa Institute for Theoretical Physics,
Kyoto University, 606-8502, Kyoto, Japan}

\author{Kazumasa Okabayashi}
\affiliation{Center for Gravitational Physics and Quantum Information, 
Yukawa Institute for Theoretical Physics,
Kyoto University, 606-8502, Kyoto, Japan}

\author{Takahiro Terada}
\affiliation{Particle Theory  and Cosmology Group, Center for Theoretical Physics of the Universe, 
Institute for Basic Science (IBS), Daejeon, 34126, Korea}

\begin{abstract}
We revisit the path-integral approach to the wave function of the Universe by utilizing Lefschetz thimble analyses and resurgence theory. 
The traditional Euclidean path-integral of gravity has the notorious ambiguity of the direction of Wick rotation. 
In contrast, the Lorentzian method can be formulated concretely with the Picard-Lefschetz theory. 
Yet, a challenge remains: the physical parameter space lies on a Stokes line, meaning that the Lefschetz-thimble structure is still unclear. Through complex deformations, we resolve this issue by uniquely identifying the thimble structure. 
This leads to the tunneling wave function, as opposed to the no-boundary wave function, offering a more rigorous proof of the previous results. 
Further exploring the parameter space, we discover rich structures: the ambiguity of the Borel resummation
of perturbative series around the tunneling 
 saddle points is exactly canceled by the ambiguity of the contributions from 
 no-boundary 
saddle points. 
This indicates that resurgence also works in quantum cosmology, particularly in the minisuperspace model. 
\end{abstract}

\date{\today}
\maketitle

\section{Introduction}

The study of quantum gravity theory remains a major challenge.
In modern approaches to quantum gravity, the gravitational path integral provides a fundamental framework for understanding the quantum behaviors of gravitational fields, and its importance is particularly emphasized in quantum cosmology. However, accurately evaluating the gravitational path integral in quantum gravity is a complex and challenging task. It faces computational obstacles, dependence on the background metric, and problems with nonperturbative effects. Despite these hurdles, the gravitational path integral remains crucial for advancing our understanding of quantum gravity \cite{Hawking:1978jz}.

We consider the gravitational path integral in general relativity, which is schematically given by 
\begin{equation}
G[g_{f};g_{i}]= \int_{{\cal M}}\mathcal{D}g_{\mu\nu}
\, \exp \left(\frac{i}{\hbar}S[g_{\mu\nu}]\right) \,.
\end{equation}
The Einstein-Hilbert action $S[g_{\mu\nu}]$
with a positive cosmological constant $\Lambda$ and a boundary term 
is written as
\begin{equation}
S[g_{\mu\nu}] = \frac{1}{2}\int_{\cal M} \mathrm{d}^4x \sqrt{-g} \left( R - 2 \Lambda\right) + \int_{\cal \partial M} \mathrm{d}^3y \sqrt{g^{(3)}} \mathcal{K}\,,
\end{equation} 
where we set $8\pi G = 1$.
The second term is the Gibbons-Hawking-York (GHY) boundary term 
with the three-dimensional metric $g^{(3)}_{ij}$ and the trace of the extrinsic
curvature $\mathcal{K}$ of the boundary $\partial {\cal M}$.
A method often utilized in the gravitational path integral is the 
formulation based on the Euclidean metric $g^{\rm E}_{\mu\nu}$.  The Hartle-Hawking no-boundary proposal~\cite{Hartle:1983ai} suggests that the wave function of the Universe is given by a path integral over compact Euclidean geometries that have a three-dimensional geometric configuration as the only boundary. This proposal elegantly explains the quantum birth of the Universe from nothing~\cite{Vilenkin:1982de} but has been criticized for various technical reasons. For instance, the path integrals over full Euclidean metrics fail to converge~\cite{Gibbons:1978ac}.
In gravity, unlike in the standard quantum field theory (QFT), these path integrals correspond to excited states related to negative energy eigenstates~\cite{Linde:1983mx}. These issues led to debates over the validity of the Euclidean gravity approach.

To avoid these problems, the authors of Ref.~\cite{Halliwell:1988ik} suggested 
 path integrals along the steepest descent paths in complex metrics. 
This method does not rely on starting with Euclidean or Lorentzian metrics. Instead, by treating it as complex, integrals are done along contours where the real part of the action decreases rapidly. However, these contours are not unique, causing ambiguity between the Hartle-Hawking no-boundary proposal~\cite{Hartle:1983ai} and Vilenkin's tunneling proposal~\cite{Vilenkin:1984wp} (related early proposals were given in Refs.~\cite{Linde:1983cm, Linde:1984ir,Rubakov:1984bh,Zeldovich:1984vk}). These proposals lead to the wave function of the Universe with the opposite exponential dependence, $G \sim \exp (\pm 12 \pi^2 / \hbar \Lambda)$
, where the $+$ and $-$ sign correspond to the no-boundary and tunneling proposal, respectively. Depending on the sign, physical consequences are drastically different~\cite{Vilenkin:1986cy,Vilenkin:1987kf}.

On the other hand, recent studies in quantum cosmology have proposed the Lorentzian path-integral formulation in a consistent way~\cite{Feldbrugge:2017kzv}. 
Integrals of phase factors such as $e^{\frac{i}{\hbar}S[g_{\mu\nu}]}$ usually do not manifestly converge, but the convergence can be ensured by shifting the contour of the integral on the complex plane by applying the Picard-Lefschetz theory~\cite{Pham,Berry:1991,Howls,Witten:2010cx}. 
\footnote{There are several studies on the quantum tunneling phenomenon, employing the Picard-Lefschetz path integral framework~\cite{Mou:2019tck,Mou:2019gyl,Millington:2020vkg,Matsui:2021oio,Rajeev:2021zae,
Hayashi:2021kro,Feldbrugge:2022idb,Nishimura:2023dky,Feldbrugge:2023frq,Feldbrugge:2023mhn}.}
According to Cauchy's theorem, the Lorentzian nature of the integral is preserved if its contour is deformed within a 
singularity-free region of the complex plane. The Lorentzian path integral can be reformulated solely in terms of the gauge-fixed lapse function, allowing direct calculation, in contrast to the Euclidean approach. 
The Lorentzian method, rooted in the Arnowitt-Deser-Misner (ADM) formalism~\cite{Arnowitt:1962hi}, is a consistent quantum gravity technique providing detailed insights into the cosmological wave function beyond general relativity~\cite{Fanaras:2021awm,Narain:2021bff,Narain:2022msz,Ailiga:2023wzl, Lehners:2023yrj,Matsui:2023hei}. 
Although the strength of the Picard-Lefschetz theory relies on the distinction between lines of the steepest descents and ascents passing through saddle points, these lines turn out to be degenerate (i.e., parameters are on a Stokes line) in the physical parameter space, obscuring the correct integration path. Therefore, more elaborate analyses are required. 
\footnote{\label{fn:contour_ambiguity} In addition, the precise integration range of the lapse function, corresponding to the interpretations of the transition amplitude as a wave function or Green's function, is still under debate~\cite{Feldbrugge:2017kzv,DiazDorronsoro:2017hti,Feldbrugge:2017mbc}.  Notably, this amounts to an ambiguity between the no-boundary wave function and the tunneling wave function for the proper wave function of the Universe in quantum cosmology. We do not solve but rather comment on this issue at the end of Sec.~\ref{sec:Lefschetz-thimble-analyses}.
}

In this paper, we provide a state-of-the-art analysis of Lorentzian quantum cosmology by utilizing Lefschetz thimble analyses and resurgence theory \cite{SC_1977__17_1_A5_0}.
The resurgence theory has a long history in applications to quantum mechanics by exact WKB analysis of the Schrodinger equation.%
\footnote{
See e.g.~Refs.~\cite{Costin:1999798,Marino:2012zq,Dorigoni:2014hea,ANICETO20191,2014arXiv1405.0356S} for review papers.
}  
Currently, there have been applications to various areas of physics from the viewpoint of not only exact WKB analyses of differential equations but also Lefschetz thimble analyses of (path) integrals. 
Recent physical applications 
include QFT, string theory~\cite{Marino:2008vx,Garoufalidis:2010ya,Chan:2010rw,Chan:2011dx,Schiappa:2013opa,Marino:2006hs,Marino:2007te,Marino:2008ya,Pasquetti:2009jg,Aniceto:2011nu,Santamaria:2013rua,Couso-Santamaria:2014iia,Grassi:2014cla,Couso-Santamaria:2015wga,Couso-Santamaria:2016vcc,Couso-Santamaria:2016vwq,Kuroki:2019ets,Kuroki:2020rgg,Dorigoni:2022bcx,Baldino:2022aqm,Schiappa:2023ned,Iwaki:2023cek,Alexandrov:2023wdj}, hydrodynamics~\cite{Aniceto:2015mto,Basar:2015ava,Casalderrey-Solana:2017zyh,Behtash:2017wqg,Heller:2018qvh,Heller:2020uuy,Aniceto:2018uik,Behtash:2020vqk}, and Jackiw-Teitelboim gravity \cite{Griguolo:2021wgy,Gregori:2021tvs,Eynard:2023qdr}. 
In particular, QFT has a variety of applications, including two-dimensional QFTs~\cite{Dunne:2012ae,Dunne:2012zk,Cherman:2013yfa,Cherman:2014ofa,Misumi:2014jua,Behtash:2015kna,Dunne:2015ywa,Buividovich:2015oju,Demulder:2016mja,Okuyama:2018clk,Marino:2019eym,Marino:2019fvu,Abbott:2020qnl,Abbott:2020mba,Marino:2021six,DiPietro:2021yxb,Marino:2021dzn,Marino:2022ykm,Reis:2022tni,Marino:2023epd}, 
Chern-Simons theory~\cite{Gukov:2016njj,Gang:2017hbs,Wu:2020dhl,Ferrari:2020avq,Gukov:2019mnk,Garoufalidis:2020nut,Fuji:2020ltq,Garoufalidis:2021osl}, the three-dimensional $O(2N)$ model \cite{Dondi:2021buw}, 
four-dimensional nonsupersymmetric QFTs~\cite{Argyres:2012vv,Dunne:2015eoa,Mera:2018qte,Canfora:2018clt,Unsal:2020yeh},
six-dimensional $\phi^3$ theory~\cite{Borinsky:2021hnd,Borinsky:2022knn}, and supersymmetric gauge theories in diverse dimensions~\cite{Russo:2012kj,Aniceto:2014hoa,Aniceto:2015rua,Honda:2016mvg,Honda:2016vmv,Honda:2017qdb,Gukov:2017kmk,Dorigoni:2017smz,Honda:2017cnz,Fujimori:2018nvz,Grassi:2019coc,Dorigoni:2019kux,Dorigoni:2021guq,Fujimori:2021oqg,Beccaria:2021ism}.
Yet, there are far fewer applications so far in the context of cosmology with some exceptions for quasinormal modes of black holes~\cite{Hatsuda:2021gtn,Hatsuda:2019eoj,Matyjasek:2019eeu,Eniceicu:2019npi} and stochastic inflation \cite{Honda:2023unh}.
This paper provides the first application of the resurgence theory to quantum cosmology.

The rest of this paper is organized as follows. In Sec.~\ref{sec:Lefschetz-thimble-analyses}, we provide a brief review of the Lorentzian quantum cosmology framework and discuss the Lefschetz thimble analyses of the gravitational path integral. 
We see that the thimble structures qualitatively change in some parameter regions as varying parameters such as the cosmological constant, boundary conditions, and so on.
We find that the thimble structure in a parameter region is ambiguous and has a discontinuity by varying the phase of $\hbar$ around ${\rm arg}(\hbar )=0$. 
In Sec.~\ref{sec:Resurgence-analyses}, we demonstrate for a technically simpler case that the ambiguity coming from the Stokes multipliers of the two no-boundary saddles is canceled from that of Borel resummation of the two tunneling saddles.
In Sec.~\ref{sec:Neumann-airy-function}, we discuss the case with the Neumann boundary condition.
Section~\ref{sec:conclusions} is devoted to conclusions.

\section{Lefschetz thimble analyses in quantum cosmology}
\label{sec:Lefschetz-thimble-analyses}

We consider the de Sitter minisuperspace model, characterized by the metric
$\mathrm{d}s^2=-\frac{N^2(t)}{q(t)} \mathrm{d}t^2+q(t)\mathrm{d}\Omega_{3}^2$, 
where $N(t)$ is the lapse function, $q(t)$ represents the scale factor squared, and $\mathrm{d}\Omega_3^2$ is the three-dimensional metric with
curvature constant $K$.
With this metric, the Einstein-Hilbert action can be expressed as
\begin{equation}\label{eq:action}
S[N,q]=V_3 \int_{0}^{1} \mathrm{d}t \left(-\frac{3\dot{q}(t)^2}{4 N(t)} +N(t)(3K-\Lambda q(t))\right)+S_B\,, 
\end{equation}
where $V_3$ denotes the three-dimensional volume factor and $S_B$ represents the potential boundary contributions at the initial $(t_i=0)$ and final $(t_f=1)$ hypersurfaces. 
Employing the Batalin-Fradkin-Vilkovisky (BFV) formalism to preserve reparametrization invariance~\cite{Fradkin:1975cq,Batalin:1977pb} and adopting the gauge-fixing condition $\dot{N}=0$, 
the gravitational transition amplitude can be written as follows:
\begin{align}\label{G-amplitude}
G\left[q(t_f);q(t_i)\right] = 
\int\! \mathrm{d}N \int \mathcal{D}q \exp\left( \frac{i S[N,q]}{\hbar} \right)\,,
\end{align}
which involves the lapse  
 integral over $N$ 
 and the path integral over all configurations of the scale factor $q(t)$~\cite{Halliwell:1988wc}.
In the above path integral, 
we need to take into account various 
boundary conditions and associated boundary terms for the background spacetime manifold.

A primary boundary condition in this framework is the Dirichlet boundary condition, which 
fixes the scale factor at two endpoints, 
$q(t_{i}=0)=q_{i}$, and $q(t_{f}=1)=q_{f}$~\cite{Feldbrugge:2017kzv}. This condition aligns with the intuitive concept of the quantum birth
of the Universe, where the Universe emerges from a state of zero size and evolves into a finite-sized entity. In this section, 
we focus on implementing this boundary condition.
Furthermore, we mainly consider the integration of the lapse function over $N \in (0, \infty)$, and this choice of $N$ ensures the causality~\cite{Teitelboim:1983fh}. One can also consider the integration on the entire real axis, $N \in (-\infty, \infty)$~\cite{DiazDorronsoro:2017hti,Feldbrugge:2017mbc}. We will briefly discuss this case at the end of Sec.~\ref{sec:Lefschetz-thimble-analyses}.

Regarding the action~\eqref{eq:action} with the Dirichlet boundary condition, the path integral 
can be exactly evaluated, and we have the following 
expression~\cite{Feldbrugge:2017kzv}:
\begin{align}\label{eq:G-amplitude-Dirichlet}
G[q_{f};q_{i}] = \sqrt{\frac{3iV_3}{4\pi\hbar}}\int_{-\infty}^\infty \! \mathrm{d}x 
\exp \left(\frac{i S_\textrm{on-shell}(x)}{\hbar}\right)\,,
\end{align}
where $N=x^2$
and the on-shell action
$S_\textrm{on-shell}(x)$ is written as
\begin{equation}\label{eq:on-shell-action}
S_\textrm{on-shell}(x) 
=\alpha x^6+ \beta x^2 + \frac{\gamma}{x^2}  \,,   
\end{equation}
with
\begin{align}
\begin{split}
&\alpha=V_3\frac{\Lambda^2}{36},\quad 
\beta =V_3\left( -\frac{\Lambda (q_i+q_f)}{2} +3K \right), \\
&\text{and} \quad \gamma = V_3\left( -\frac{3}{4} (q_f-q_i)^2\right)\,. 
\end{split}
\end{align}
We have changed the integration range of $x$ from $(0, \infty)$ to the sum of $(-\infty, 0)$ and $(0, \infty)$ and divided it by $2$.

Now, we consider the perturbative expansion of $\hbar$ around saddle points in the gravitational path integral~\eqref{eq:G-amplitude-Dirichlet}, and rewrite the integral in terms of steepest descents associated with contributing saddles. 
For this purpose, it is convenient to work in the following expression:
\begin{equation}
G(\hbar) 
= \sqrt{\frac{3iV_3}{4\pi\hbar}}\int_{-\infty}^\infty  \! \mathrm{d}x \exp [F(x)] \,,
\label{eq:integralF}
\end{equation}
where we introduced 
\begin{equation}\label{eq:exponential-part1}
F(x) := \frac{i}{\hbar} S_\textrm{on-shell}(x) \,,
\end{equation}
as the exponential part of the above integration.

We study the behavior of this integration by using the saddle-point method. 
The derivative of the exponential part $\frac{\mathrm{d}F(x)}{\mathrm{d}x}=0$
leads to the eight saddle points, 
\begin{align}
x_n^\pm 
=e^{\frac{\pi i}{2}n } \left( \frac{ \pm (-1)^n \sqrt{12 \alpha  \gamma  +\beta ^2}
-\beta }{6\alpha } \right)^{1/4} , 
\label{eq:saddles}
\end{align}
where $n=0, 1, 2,$ and $3$. 
It is found that $x_n^+$ and $x_n^-$
are the tunneling and no-boundary saddle points, respectively. 
For the corresponding saddle points, we have 
\begin{align}
F(x_n^\pm )
=\frac{(-1)^n i \sqrt{\frac{2}{3}} \left(12 \alpha  \gamma -\beta ^2 \pm (-1)^n\beta  \sqrt{12 \alpha  \gamma +\beta ^2}\right)}
 {3 \alpha  \hbar \sqrt{\frac{\pm (-1)^n \sqrt{12 \alpha  \gamma +\beta ^2}-\beta }{\alpha  }}}. 
\end{align}
In the simplest model to describe the quantum creation of the Universe from nothing, by definition, we can take $\alpha > 0$, $\gamma < 0$, and $\beta$ as either positive or negative. 
By using the saddle-point method, for $\textrm{Re}[F(x_n^+)]<0$ and $\textrm{Re}[F(x_n^- )]>0$, 
we obtain the tunneling and no-boundary wave functions, respectively.

Note that, in general, the saddle points not on the original integral contour may or may not contribute and we need further analysis to determine which saddles contribute.
The standard way to see this is to find the steepest descents or Lefschetz thimbles associated with the saddle points and then deform the integral contour to an appropriate superposition of the thimbles equivalent to the original one. 
The Lefschetz thimble $\mathcal{J}_{x_s}$ associated with the saddle point $x_s$ has the following important properties: (a) $\textrm{Im}[F(x)] =\textrm{Im}[F(x_s)]$ along $x \in \mathcal{J}_{x_s}$ and 
(b) $\textrm{Re}[F(x)]$ monotonically decreases as we go far away from $x_s$ along $x \in \mathcal{J}_{x_s}$.
Then we rewrite the integral as\footnote{
Because of the property (b), Lefschetz thimbles connect regions where the integrand vanishes. Therefore, to rewrite the integral as a superposition of the Lefschetz thimbles, the original integral contour should be closed or have a vanishing integrand at its endpoints.
}
\begin{equation}
G(\hbar)=\sqrt{\frac{3iV_3}{4\pi\hbar}} \sum_{x_s} n_{x_s} \int_{\mathcal{J}_{x_s}} \mathrm{d} x \exp [F(x)] \,, 
\label{eq:decomposition}
\end{equation}
where $n_{x_s}$ is an integer called the Stokes multiplier to determine how the saddle $x_s$ contributes.

In general, as the parameters are changed, the integer $n_{x_s}$ cannot change continuously but may have jumps across particular regions of the parameters called Stokes lines.
In that case, the $\hbar$ expansion changes its form by varying the parameters, and this is called the Stokes phenomenon.
The Stokes phenomenon can occur when there are multiple saddle points with the same imaginary part of the exponent in the integrand, i.e.,
\begin{equation}
\textrm{Im}[F(x_s)]=\textrm{Im}[F(x_s')]\,,
\end{equation}
where $x_s$ and $x_s'$ denote different saddle points. 
On the Stokes lines, the Lefschetz thimbles pass multiple saddle points and the Stokes multiplier $n_{x_s}$ is not unique.
Therefore, the thimble decomposition \eqref{eq:decomposition} is ambiguous on the Stokes lines.
In this situation, it is often convenient to go slightly away from the Stokes lines by changing the parameters and seeing what happens as we approach the Stokes lines from different directions.
Indeed, there are various examples where the thimble decomposition includes terms having jumps across the Stokes lines, but the total answer does not have the jumps due to the cancellation of the ambiguity against other ambiguities coming from the subtlety of resummation in perturbative series.

\begin{figure}[t] 
\includegraphics[width=0.48\textwidth]{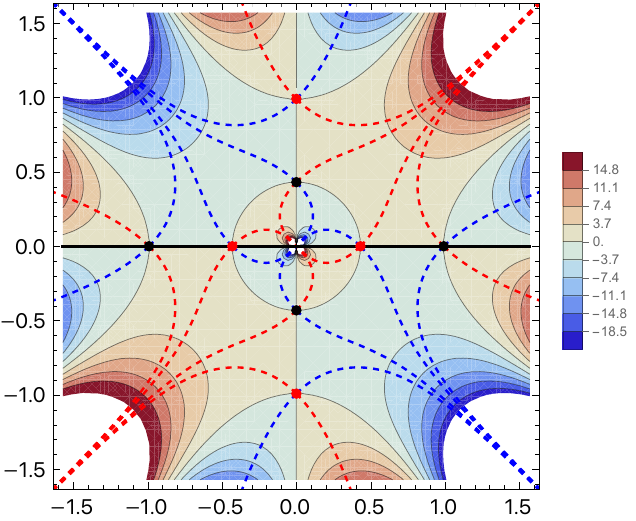}
	\caption{
 Contour plot of $\textrm{Re}\left[F(x)\right]$ 
 over the complex $x$ plane for $\alpha =1 $, $\beta =-3 $, $\gamma =-0.1 $, and $\hbar =1$, corresponding to the parameter region (i) $12\alpha \gamma +\beta^2 >0$ with $\beta <0$. 
 The black (red) circles represent the tunneling (no-boundary) saddle points, while the blue (red) lines denote the (dual) Lefschetz thimbles. The black horizontal line denotes the original integration contour.
 }
	\label{fig:Picard-Lefschetz1}
\end{figure} 

Let us study the thimble structures of the integral \eqref{eq:integralF}.
As we will see soon, the structures are qualitatively different among the three parameter regions: 
\begin{enumerate}
\renewcommand{\theenumi}{(\roman{enumi})}
\item $12\alpha \gamma +\beta^2 >0$ with $\beta <0$\,,
\item $12\alpha \gamma +\beta^2 <0$\,,
\item  $12\alpha \gamma +\beta^2 >0$ with $\beta >0$\,,
\end{enumerate}
where 
$12\alpha\gamma+\beta^2=V_3^2\left [q_fq_i\Lambda^2-3K\Lambda(q_i+q_f)+9K^2\right ]$.
We comment on the physical meanings of these parameter regions. In quantum cosmology, the spatial manifold is generally assumed to be closed; therefore, our discussion mainly focuses on a closed universe with $K>0$ and a positive cosmological constant $\Lambda>0$. Region (i) corresponds to a transition from the initial Lorentzian state ($q_i > 3K/\Lambda$) to the final Lorentzian state ($q_f > 3 K / \Lambda$).  Region (ii) corresponds to a transition involving one Lorentzian state and one Euclidean state [$(q_i-3K/\Lambda)(q_f - 3K/\Lambda)< 0$].  Region (iii) corresponds to a transition from the initial Euclidean state ($q_i < 3 K / \Lambda$) to the final Euclidean state ($q_f < 3 K /\Lambda$).  In quantum cosmology, the initial state representing nothing ($q_i = 0$) is often considered, which must be Euclidean because of the positive spatial curvature. In this case, region (i) cannot be realized, while regions (ii) and (iii) correspond to the final Lorentzian and Euclidean spacetime, respectively.

Figure~\ref{fig:Picard-Lefschetz1} shows the contour plot of $\textrm{Re}\left[F(x)\right]$ in Eq.~\eqref{eq:exponential-part1} over the complex $x$ plane for $\alpha =1$, $\beta =-3$, $\gamma =-0.1$, and $\hbar =1$, which are representative values of the parameters in region (i).
Note that the absolute value of $\hbar$ does not affect the thimble structures since it just changes the overall scaling of $F(x)$, while the phase of $\hbar$ turns out to be important, as we will see later.
From Fig.~\ref{fig:Picard-Lefschetz1}, we see that the four saddle points on the real axis contribute and the integral can be rewritten as a superposition of their thimbles unambiguously.
Therefore, we can clearly say which saddles contribute in the case of the region (i), while we will see that the other cases are more intricate. 

\begin{figure}[t] 
\includegraphics[width=0.48\textwidth]{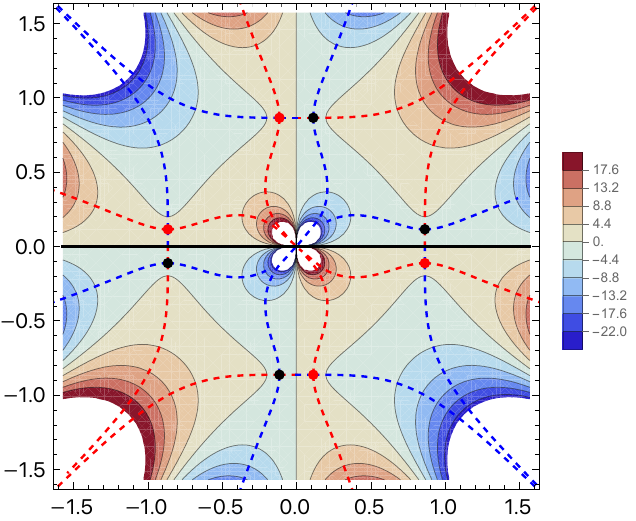}
	\caption{
 A similar plot to Fig.~\ref{fig:Picard-Lefschetz1} with $\alpha =1$, $\beta =-3$, $\gamma =-1$ corresponding to the parameter region (ii), $12\alpha \gamma +\beta^2 <0$. This corresponds to the no-boundary and tunneling proposals.}
	\label{fig:Picard-Lefschetz2}
\end{figure} 
\begin{figure}[t] 
\subfigure[$\theta =+\frac{\pi }{10}$]{%
\includegraphics[width=0.48\textwidth]{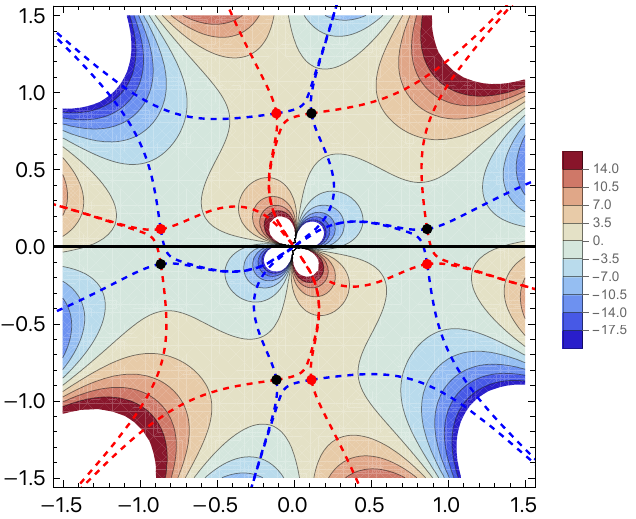}}\\
\subfigure[$\theta =-\frac{\pi }{10}$]{%
\includegraphics[width=0.48\textwidth]{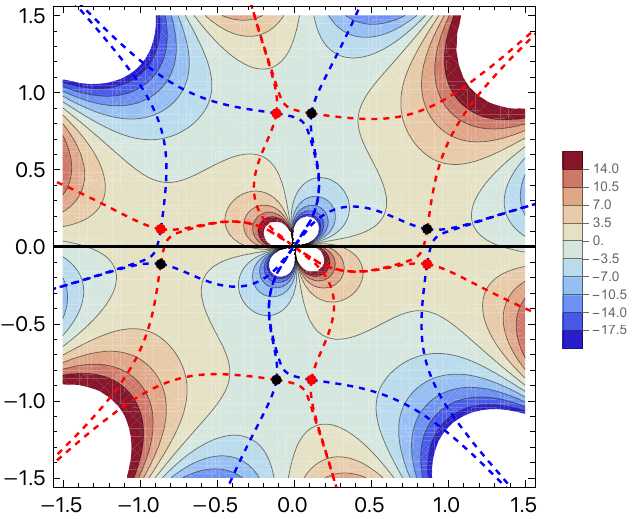}}
	\caption{
 Complex deformation of Fig.~\ref{fig:Picard-Lefschetz2} [in region (ii)] with $\hbar = e^{i\theta }$.
 }
	\label{fig:Picard-Lefschetz2-2}
\end{figure} 

In Figure~\ref{fig:Picard-Lefschetz2}, we show a similar plot to Fig.~\ref{fig:Picard-Lefschetz1} in the parameter region (ii). 
A main difference from Fig.~\ref{fig:Picard-Lefschetz1} is that some Lefschetz thimbles are passing two saddle points.
This implies that we are on the Stokes lines 
and the decomposition \eqref{eq:decomposition} is ambiguous at the values of the parameters. 
In particular, we cannot judge how the no-boundary and tunneling saddle points contribute to the integral \eqref{eq:integralF} just by Fig.~\ref{fig:Picard-Lefschetz2}.
To see what is happening more precisely, we consider a deformation of the setup to go away from the Stokes lines.
In particular, we take the expansion parameter $\hbar$ to be complex, $e^{i\theta}$, as is usually done in the context of resurgence.
Figure~\ref{fig:Picard-Lefschetz2-2} shows similar plots to Fig.~\ref{fig:Picard-Lefschetz2} but with nonzero $\theta$. \footnote{To ensure the convergence at the origin and at infinity, we tilt the direction of the integration contour by an angle $\theta_x$ satisfying, $\theta_x > \max(\theta/6, - \theta/2)$. For example, we may take $\theta_x = |\theta|$ for a sufficiently small $|\theta|$. Then, the contours reduce to the original ones as $\theta \to 0$. A similar comment applies to the parameter region (iii).}
Now we find that all of the Lefschetz thimbles cross only one saddle point, in contrast to the $\theta =0$ case in Fig.~\ref{fig:Picard-Lefschetz2}.
Furthermore, the contributing saddle points are the same between $\theta >0$ and $\theta <0$: only the two tunneling saddles around the real axis contribute.
This structure is maintained as long as $\theta$ is not strictly zero.
The $\theta =0$ case of interest can be understood as the limit $\theta \rightarrow 0^\pm$ from nonzero $\theta$.
The plots in Fig.~\ref{fig:Picard-Lefschetz2} imply that the limit is smooth and independent of the directions of the limit.\footnote{
A similar thing happens, e.g., when we consider an integral of $\exp \left[ -\frac{1}{\hbar} (x^2 +1)^2  \right]$ along $x\in \mathbb{R}$. In this example, the Lefschetz thimble associated with the saddle $x=\pm i$ crosses the trivial saddle $x=0$ for $\hbar\in\mathbb{R}$. In contrast, the thimble decomposition is unique for nonzero $\theta$ and the contributing saddle (i.e.,~$x=0$) is the same between $\theta >0$ and $\theta <0$.
}
Therefore, we conclude that the contributing saddles in the parameter region (ii) are the two tunneling saddles around the real axis.
Note that while the same conclusion was made in \cite{Feldbrugge:2017kzv}, it cannot be justified by resolving the thimble structures on the Stokes lines in only one direction of the deformation parameter. Our analysis justifies the conclusion by going off the Stokes line in multiple directions and then taking the limits to the original setup.
Our finding conclusively demonstrates that in traditional quantum cosmology, the tunneling proposal, rather than the no-boundary proposal, is the only approach that accurately describes the quantum creation of the Universe from nothing.

\begin{figure}[t] 
\includegraphics[width=0.48\textwidth]{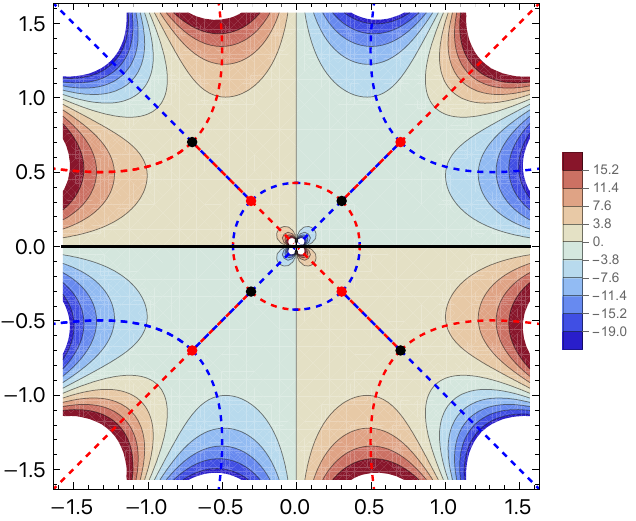}
	\caption{
 A similar plot to Figs.~\ref{fig:Picard-Lefschetz1} and \ref{fig:Picard-Lefschetz2} with $\alpha =1 $, $\beta =3 $ and $\gamma =-0.1 $ corresponding to region (iii), $12\alpha \gamma +\beta^2 >0$, with $\beta >0$. }
	\label{fig:Picard-Lefschetz3}
\end{figure} 
\begin{figure}[t] 
\subfigure[$\theta =+\frac{\pi }{10}$]{%
\includegraphics[width=0.48\textwidth]{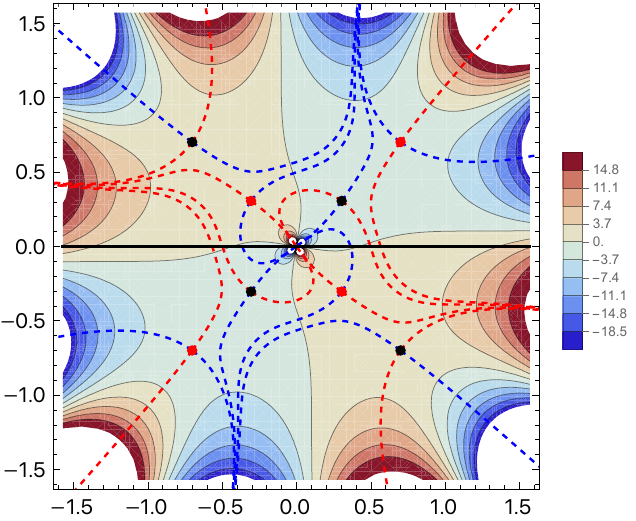}}\\
\subfigure[$\theta =-\frac{\pi }{10}$]{%
\includegraphics[width=0.48\textwidth]{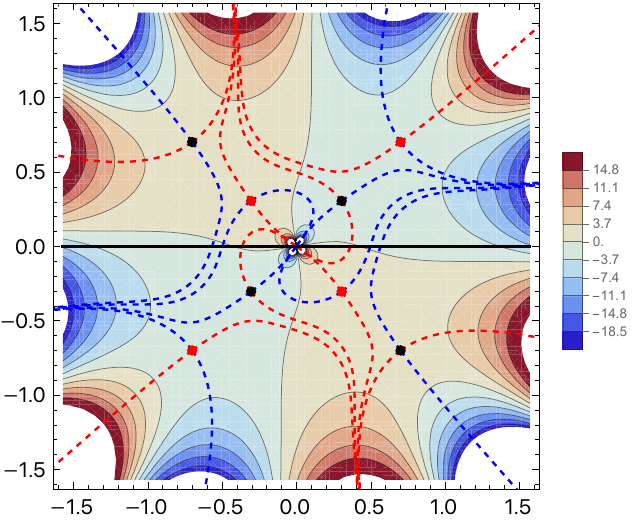}}
	\caption{
 Complex deformation of Fig.~\ref{fig:Picard-Lefschetz3} [in region (iii)] with $\hbar = e^{i\theta }$.
 }
	\label{fig:Picard-Lefschetz3-2}
\end{figure} 

Let us turn to the parameter region (iii), whose thimble structure for $\theta =0$ is presented in Fig.~\ref{fig:Picard-Lefschetz3}.
Again, we see that some Lefschetz thimbles pass two saddle points and therefore we are on the Stokes lines. 
To resolve this, 
let us turn on the nonzero $\theta$ as in the region (ii) case.
Figure~\ref{fig:Picard-Lefschetz3-2} shows similar plots to Fig.~\ref{fig:Picard-Lefschetz3} for nonzero $\theta$, which are also similar to Fig.~\ref{fig:Picard-Lefschetz2-2} but with a different parameter region. 
Here, all of the Lefschetz thimbles cross only one saddle point, as in the parameter region (ii) in Fig.~\ref{fig:Picard-Lefschetz2-2}.
A main difference is that the contributing saddles are {\it different} between $\theta >0$ and $\theta <0$: all four saddles in the first and third quadrants contribute for $\theta >0$, while only the two tunneling saddles in the quadrants contribute for $\theta <0$.
This structure is maintained as long as $\theta$ is not strictly zero.
This implies that the Stokes multipliers $n_{x_s}$ for the two no-boundary saddles in the quadrants introduced in Eq.~\eqref{eq:decomposition} are like step functions jumping at $\theta =0$!
Does this mean that the original integral itself \eqref{eq:integralF} is discontinuous at $\theta =0$?
We will see that the answer is no: although each term of the summand in Eq.~\eqref{eq:decomposition} may be ambiguous due to a jump at $\theta =0$, the ambiguity is canceled by that coming from the other terms and then the whole answer is continuous.
In other words, the limits $\theta\rightarrow 0^+$ and $\theta\rightarrow 0^-$ can be different for each term but are the same for the whole answer.
It is known that this kind of cancellation typically happens in successful examples of the resurgence where ambiguities for Stokes multipliers of some saddles are canceled by ambiguities coming from the resummation of perturbative series around other saddles.
In the next section, we explicitly see the cancellations among the ambiguities coming from the Stokes multipliers of the two no-boundary saddles and the Borel resummation of the two tunneling saddles.

\subsection{Comments on alternative integration range of lapse function }

So far, we have considered the integration of the lapse function over $N \in (0, \infty)$ and 
this choice of $N$ ensures the causality~\cite{Teitelboim:1983fh}.
On the other hand, 
considering all ranges of the lapse function $N \in (-\infty, \infty)$ is also possible~\cite{DiazDorronsoro:2017hti,Feldbrugge:2017mbc}, which satisfies the gauge invariance of the lapse function. 
In this subsection, we comment on this case with all integration ranges of the lapse function $N \in (-\infty, \infty)$. 
Here, as in the previous section where we changed the integration range of $x$ from $(0, \infty)$ to the sum of $(-\infty, 0)$ and $(0, \infty)$ and then divide it by $2$, we similarly change the integration range of $x$ after substituting $N=x^2$.

The authors of Refs.~\cite{DiazDorronsoro:2017hti,Feldbrugge:2017mbc} considered the deformation of the integration contour along the entire real axis to avoid singularities at the origin. 
In particular, they advocated the deformation into the lower half-plane of $N$ to obtain the Hartle-Hawking no-boundary wave function.
Let us compare their results with our discussion in terms of $x$ based on the resurgence technique.  
The integrand of $G(\hbar)$ (defined with the alternative contour) has a factor of $1/\sqrt{N}$ having a branch-cut singularity.%
\footnote{Lefschetz thimble structures of integrals with branch cuts were studied in, e.g.,~Refs.~\cite{Kanazawa:2014qma,Fujii:2015vha,Fujimori:2018nvz,Fujimori:2021oqg}.} 
In this subsection, we take $\text{arg}(N) \in (-\pi, \pi)$. When the contour of $N$ passes above the origin, there are two contours in terms of $x$. One is the path in the first quadrant asymptoting in the directions of the positive imaginary axis and the positive real axis. The second one is the path of $x$ in the third quadrant asymptoting in the directions of the negative real axis and the negative imaginary axis. On the other hand, when the contour of $N$ passes below the origin, there are also two contours in terms of $x$. One is the path asymptoting in the directions of the negative imaginary axis and the positive real axis. The second one is the path asymptoting in the directions of the negative real axis and the positive imaginary axis.

In the parameter region (i), the contour of $N$ avoiding the origin on the upper half-plane is equivalent to the contour of $x$ composed of Lefschetz thimbles associated with both the no-boundary and tunneling saddle points on the imaginary axis and those on the real axis. 
The contour of $N$ avoiding the origin on the lower half-plane is equivalent to the contour of $x$ composed of Lefschetz thimbles associated with 
the no-boundary saddle points on the imaginary axis and the tunneling saddle points on the real axis.  

The parameter region (ii) is on the Stokes lines, as we saw above. 
For the contour of $N$ avoiding the origin on the upper half-plane, the contributing saddle points in terms of $x$ are the tunneling saddle points in the first and third quadrants for both signs of the deformation angle $\theta$.  
For the contour of $N$ avoiding the origin on the lower half-plane, however, the contributing saddle points depend on the sign of the deformation angle $\theta$, in contrast to the analysis for only positive $N$. 
In particular, different tunneling saddles contribute for different signs of $\theta$ while the no-boundary saddles contribute for both signs as seen from Fig.~\ref{fig:Picard-Lefschetz2-2}. 
Given this situation, it is unclear whether or not the $\theta \rightarrow 0^\pm$ limit of $G(\hbar )$ is real and satisfies the Wheeler-DeWitt equation.
More detailed analysis is needed to have a clear answer.

The parameter region (iii) is also on the Stokes lines. 
For the contour of $N$ avoiding the origin on the upper half-plane, the contributing saddle points in terms of $x$ are the no-boundary saddle points in the first and third quadrants for both signs of $\theta$. 
This is in contrast to the Stokes phenomenon found in the parameter region (iii) in the case of positive $N$.  
For the contour of $N$ avoiding the origin on the lower half-plane, 
the contributing saddle points depend on the deformation angle $\theta$, signaling the Stokes phenomenon. See Fig.~\ref{fig:Picard-Lefschetz3-2}. 

In this way, the ambiguity dependent on the original definition of the integration contour of $N$ remains as we work with $x$. This is not due to the insufficient power of our resurgence analyses, but rather the different definitions of the original contour of $N$ simply correspond to inequivalent quantities.

\section{Resurgence analyses in Lorentzian quantum cosmology }
\label{sec:Resurgence-analyses}

In this section, we apply the resurgence method to the Lorentzian path integral.
For simplicity, we consider the persistence of spacetime where $q_i=q_f$ and the size of the Universe does not change over time.
In this case, we have $\gamma =0$ and obtain the following expression:
\begin{equation}\label{eq:G-amplitude-persistence}
G(\hbar)=\sqrt{\frac{3iV_3}{4\pi}}
\int^{\infty}_{-\infty} \mathrm{d} x \exp \left[i \left(\alpha\hbar^2x^6 +\beta x^2 \right) \right] \,,
\end{equation}
where we transform $x\to x\sqrt{\hbar}$ and consider the integration range of $x$ from $-\infty$ to $+\infty$, 
for simplicity. 
There are five saddle points of the exponent,
\begin{equation}
    x=0,\quad x_{i}= e^{\frac{i \arg \beta}{4}} e^{\frac{2n-1}{4} \pi i} \left(\frac{|\beta|}{3 \alpha \hbar^2 }\right)^{\frac{1}{4}} \quad (n=1,...,4),
\end{equation}
where we take $\arg \beta=-\pi$ for $\beta<0$ such that $0 \le \arg{x_i} <2\pi$.
From the viewpoint of the $\gamma \neq 0$ case, the trivial saddle $x=0$ comes from degeneration of four saddles out of the eight saddles in Eq.~\eqref{eq:saddles}, which are $x_n^+$ with even (odd) $n$ and $x_n^-$ with odd (even) $n$ for $\beta >0$ ($\beta < 0$).

Let us consider deriving the exact result from the perturbation expansion from the saddle points.
First, we consider the contribution of the trivial saddle point, $x=0$,
\begin{equation}
\sqrt{\frac{3iV_3}{4\pi }} \int_{\mathcal{J}_0} \mathrm{d} x \exp \left[ F\left( \sqrt{\hbar} x \right) \right] ,
\end{equation}
and its small-$\hbar$ expansion formally given by
\begin{align}\label{eq:asymptotic_series_trivial}
    G_{0}(\hbar) = \sum_{n=0}^\infty c_n \hbar^{2n} \, .
\end{align}
To compute the perturbative coefficient $c_n$, we can replace the integral contour $\mathcal{J}_0$ with $(-e^{\pm \pi i/4} \infty ,e^{\pm \pi i/4} \infty )$ for ${\rm sign}(\beta)=\pm 1$.
Then the problem is reduced to integrals of polynomials of $x$ with the Gaussian weight and the coefficient $c_n$ is given by
\begin{equation}
    c_n = e^{ \pm \frac{\pi i}{4}} \sqrt{\frac{3 i V_3 }{4\pi |\beta|}} \cdot 
        \frac{\Gamma(3n+\frac{1}{2})}{\Gamma(n+1)} 
        \left(\frac{\alpha}{\beta^3}\right)^n,
\end{equation}
according to ${\rm sign}(\beta)=\pm 1$.
As a result of the formal $\hbar$ expansion, it is found that Eq.~(\ref{eq:asymptotic_series_trivial}) is an asymptotic series. Hence, we take its Borel resummation. As a first step of the resummation, we consider the Borel transformation ${\cal B}G_0(t)$ of Eq.~(\ref{eq:asymptotic_series_trivial}). 
It is given by
\begin{equation}\label{BorelSeries_trivial}
    {\cal B}G_0(t)=
    \sum_{n=0}^\infty \frac{c_n}{\Gamma(2n+1)}t^{2n}\,.
\end{equation}
This is a convergent series, and the infinite sum has an analytical expression for ${\rm sign}(\beta)=\pm 1$:
\begin{align}\label{eq:BorelTrans_trivial}
    {\cal B}G_0(t)=
        e^{\pm \frac{i \pi}{4}} \sqrt{\frac{3 i V_3}{4 |\beta|}} ~ 
        _2F_1\left(\frac{1}{6},\frac{5}{6};1;\frac{27 t^2 \alpha }{4 \beta ^3}\right) \,,
\end{align}
where 
$_2F_1(a, b; c; z)$ is a hypergeometric function.
We denote the Borel resummation of Eq.~(\ref{eq:G-amplitude-persistence}) as $\mathcal{S}_{\theta}G_0(\hbar)$, and it is defined by the Laplace transformation of Eq.~(\ref{eq:BorelTrans_trivial}):
\begin{equation} 
    \mathcal{S}_{\theta}G_0(\hbar)= 
    \frac{1}{\hbar} \int_{0}^{\infty\cdot e^{i \theta}}\mathrm{d} t \, e^{-\frac{t}{\hbar}} {\cal B}G_0(t)\,,
    \label{eq:Borel0}
\end{equation}
where $\theta$ is the argument of $\hbar$.

For $\alpha>0$ and $\beta <0$, the Borel resummation $\mathcal{S}_{\theta}G_0(\hbar)$ is well defined along $\theta=0$ and given by
\begin{align}
\begin{split}
\mathcal{S}_{0}G_0(\hbar) 
    &= 
    \sqrt{\frac{3 
    V_3}{4}} \cdot \frac{\pi}{(12 \alpha \hbar^2)^{1/6}} \times \\
    & \quad \left[\text{Ai}\left(\frac{\beta}{(12 \alpha \hbar^2)^{1/3} }\right)^2+\text{Bi}\left(\frac{\beta}{(12 \alpha \hbar^2)^{1/3} }\right)^2\right]\,,
\end{split}
\end{align}
where $\text{Ai}(x)$ and $\text{Bi}(x)$ are Airy functions of the first and second kind, respectively.\footnote{
The contributions of the other saddles are 
\begin{align*}
   & G(\hbar)|_{x=x_1} =  G(\hbar)|_{x=x_3} 
   \nonumber \\
    =& (-1)^{1/3} \frac{\pi^{3/2}\sqrt{V_3}}{2\alpha^{1/6}\hbar^{1/3}} G_{5,3}^{0,3}\left( \frac{19683 \alpha^3 \hbar^6}{\beta^9} , 3 \left| \begin{matrix} \frac{1}{3}, \frac{2}{3}, 1, \frac{1}{2}, \frac{5}{6}  \\ \frac{1}{6}, \frac{1}{2}, \frac{5}{6} \end{matrix} \right. \right) \nonumber \\
    & -  \frac{
    \pi \sqrt{3 
    V_3}}{4 (12 \alpha \hbar^2)^{\frac{1}{6}}} \left[\text{Ai}\left(\frac{\beta}{(12 \alpha \hbar^2)^{\frac{1}{3}} }\right)^2+\text{Bi}\left(\frac{\beta}{(12 \alpha \hbar^2)^{\frac{1}{3}} }\right)^2\right]\,,
\end{align*}
where $G_{p,q}^{m,n}\left(z, r \left| \begin{matrix} a_1, \dots, a_p \\ b_1, \dots, b_q \end{matrix} \right. \right)$ is the generalized Meijer G-function. 
The total amplitude is 
\begin{align*}
    G(\hbar)= (-1)^{1/3} \frac{\pi^{3/2}\sqrt{V_3}}{\alpha^{1/6}\hbar^{1/3}} G_{5,3}^{0,3}\left( \frac{19683 \alpha^3 \hbar^6}{\beta^9} , 3 \left| \begin{matrix} \frac{1}{3}, \frac{2}{3}, 1, \frac{1}{2}, \frac{5}{6}  \\ \frac{1}{6}, \frac{1}{2}, \frac{5}{6} \end{matrix} \right. \right).
\end{align*}
}

For $\alpha>0$ and $\beta >0$, the Borel resummation $\mathcal{S}_{\theta}G_0(\hbar)$ along $\theta=0$ is not well defined and non-Borel summable due to the branch cut of the hypergeometric function. 
One may try to make it well defined by deforming the integral contour in Eq.~\eqref{eq:Borel0} to avoid the branch cut, but then the value of the integral depends on how to avoid the branch cut.   
This ambiguity is estimated by the difference between the Borel resummations along the directions $\theta =0^+$ and $\theta =0^-$ as
\begin{align}\label{eq:Borel_ambiguity}
\begin{split}
&\left(\mathcal{S}_{0^{+}}-\mathcal{S}_{0^{-}} \right)G_0(\hbar)
    = \lim_{\epsilon \to 0} \frac{i 
    } {\hbar} \sqrt{\frac{3 
    V_3}{4 \beta}} ~
    \int_{t_0}^{\infty} \mathrm{d}t \, e^{-\frac{t}{\hbar}}\times \\
& \left[ {}_2F_1\left(\frac{1}{6},\frac{5}{6};1;\frac{27 t^2 \alpha }{4 \beta ^3}
    +i\epsilon\right)- {}_2F_1\left(\frac{1}{6},\frac{5}{6};1;\frac{27 t^2 \alpha }{4 \beta ^3}
    -i\epsilon\right) \right]\\
&=-\frac{1 
} {\hbar} \sqrt{\frac{3 
V_3}{4 \beta}}
    \int_{t_0}^{\infty} \mathrm{d}t \, e^{-\frac{t}{\hbar}} {}_2F_1\left(\frac{1}{6},\frac{5}{6};1;1-\frac{27 t^2 \alpha }{4 \beta ^3}\right)\,,
\end{split}
\end{align}
where $t_0=(4\beta^3/27\alpha)^{1/2}$. 
As seen later, this ambiguity cancels the ambiguity of the Stokes multipliers for the no-boundary saddle points.
To see the cancellation, we integrate Eq. (\ref{eq:Borel_ambiguity}) after $\hbar$ expansion of the integrand and obtain
\begin{align}\label{eq:Borel_ambiguity_perturbation}
\begin{split}
    &\left(\mathcal{S}_{0^{+}}-\mathcal{S}_{0^{-}} \right)G_0(\hbar)\\
    &= - 
    e^{ -\frac{t_0 }{\hbar} } \sqrt{\frac{3 
    V_3}{4 \beta}}
    \left[ 1- \frac{5}{12}\left( \frac{3\alpha}{\beta^3} \right)^{1/2}\hbar + \mathcal{O}(\hbar^2) \right].
\end{split} 
\end{align}
Next, we consider the contribution of the nontrivial saddle points, $x=x_1$ and $x_3$ for $\alpha>0$ and $\beta>0$.
In this case, the angles of the steepest descent contours at the nontrivial saddle points are both $-\pi/4$. 
Hence, the contribution of the thimble associated with $x=x_1$ to the integral Eq.~(\ref{eq:G-amplitude-persistence}) is rewritten as 
\begin{align}
G(\hbar )|_{x=x_1} 
&:=n_{x_1} \sqrt{\frac{3iV_3}{4\pi }} \int_{\mathcal{J}_{x_1}} \mathrm{d} x \exp \left[ F\left( \sqrt{\hbar} x \right) \right] \nonumber \\
&= n_{x_1} 
e^{-\frac{t_0}{\hbar}} \sqrt{\frac{3 
V_3}{4 \beta}} 
        \int^{\infty}_{-\infty} {\rm d}\rho~ e^{-4\beta \rho^2} e^{i \sigma_1(\rho)},
\end{align}
where we have substituted $x=x_1 + \rho e^{-\pi i /4}$ and
\begin{align}\label{eq:integral_x1}
\begin{split}
    \sigma_1(\rho) = 
    &-20\left( \frac{\alpha \beta^3} {27} \right)^{1/4}\hbar^{1/2} \rho^3- 5 i (3\alpha \beta)^{1/2}\hbar \rho^4 \\
    & \quad + 2 (27 \alpha^3 \beta)^{1/4} \hbar^{3/2} \rho^5 + i \alpha \hbar^2 \rho^6 .
\end{split}
\end{align}
Integrating after $\hbar$ expansion of the integrand, we obtain 
\begin{align}\label{eq:contribution_x1}
\begin{split}
    &G(\hbar)|_{x=x_1} \\
    & = \frac{n_{x_1}}{2} 
    e^{ -\frac{t_0 }{\hbar} } \sqrt{\frac{3 
    V_3}{4 \beta}}
    \left[ 1- \frac{5}{12}\left( \frac{3\alpha}{\beta^3} \right)^{1/2}\hbar + \mathcal{O}(\hbar^2) \right].
\end{split}
\end{align}
Since the contribution from the thimble passing through the saddle point $x=x_3$ is the same as that for $x=x_1$, we find the cancellation of the Borel ambiguity by the ambiguity of the Stokes multipliers for the nontrivial saddle points,
\begin{align}\label{eq:cancellation_of_ambiguity}
&    \left(\mathcal{S}_{0^{+}} \! -\mathcal{S}_{0^{-}} \right)G_0(\hbar) 
= - \!\!\! \!\! \!  \sum_{x_s =x_1 , x_3 } \! \! \! \! \! \left ( G(\hbar)|_{x=x_s }^{\theta =0^+} \!  -G(\hbar)|_{x=x_s}^{\theta =0^- }     \right ) ,
\end{align}
at the level of $\hbar$ expansion. We have analytically confirmed this cancellation up to the $100$th order of $\hbar$.\footnote{
We thank Kei-ichiro Kubota for efficient implementation ideas for the calculation. 
}
 In other words, the total $G(\hbar)$ in Eq.~\eqref{eq:G-amplitude-persistence} is well defined and continuous at $\theta = 0$.\footnote{
The total contribution of the amplitude is
\begin{align*}
    G(\hbar) = (-1)^{1/3} \frac{\sqrt{V_3}}{4 \pi^{3/2} \alpha^{1/6}\hbar^{1/3}} G_{4,2}^{1,3} \left( - \frac{19683 \alpha^3 \hbar^6}{\beta^9}, 3 \left| \begin{matrix} \frac{1}{3}, \frac{2}{3}, 1 \\ \frac{1}{6} \end{matrix} \right. \right).
\end{align*}
}

\section{Neumann boundary condition and airy function}
\label{sec:Neumann-airy-function}

In the previous section, we considered the Dirichlet boundary condition, which is a common choice in quantum cosmology. However, it is possible to consider other, nontrivial boundary conditions for the background geometry such as Neumann and Robin boundary conditions~\cite{DiTucci:2019dji,DiTucci:2019bui,Narain:2021bff,
Narain:2022msz,Ailiga:2023wzl}. 
We note that it is found that the tunneling saddle points lead to unstable perturbations and the Dirichlet boundary condition raises challenges in Lorentzian quantum cosmology~\cite{Feldbrugge:2017fcc}. Specifically, adopting the Dirichlet boundary condition in the parameter region of no-boundary and tunneling proposals, the Lefschetz thimbles fail to avoid these unstable saddle points; hence, this boundary condition leads to the issue of perturbation instability. Despite extensive discussion and numerous studies addressing this problem~\cite{ DiazDorronsoro:2017hti, Feldbrugge:2017mbc, Feldbrugge:2018gin, DiazDorronsoro:2018wro,Halliwell:2018ejl,Janssen:2019sex, Vilenkin:2018dch, Vilenkin:2018oja, Bojowald:2018gdt, DiTucci:2018fdg, DiTucci:2019dji, DiTucci:2019bui, Lehners:2021jmv,Matsui:2022lfj,Matsui:2024bfn}, 
adopting nontrivial and imaginary boundary conditions is a straightforward solution~\cite{DiTucci:2019dji,DiTucci:2019bui}.

We consider the Neumann boundary condition.  In general, it fixes 
$\dot{q}\mid_{t=t_i}$ and/or $\dot{q}\mid_{t=t_f}$,
and the corresponding GHY term should be removed at $t=t_i$ and/or $t=t_f$. 
Since it has been pointed out that the Euclidean initial momentum is crucial for the success of the no-boundary proposal~\cite{DiTucci:2019dji,DiTucci:2019bui}, we impose the Neumann boundary condition at the initial time $t=t_i = 0$. 
Imposing the Euclidean (imaginary) Neumann boundary conditions on the initial hypersurface, the perturbative instability can be avoided. Furthermore, the ambiguity of the contour of complex metrics can be removed.

Hereafter, we consider the case of
imposing the Neumann boundary condition on the initial hypersurface and the Dirichlet boundary condition on the final hypersurface, 
\begin{equation}\label{eq:Dirichlet-boundary}
-\frac{3\dot{q}(t_{i}=0)}{2N}=p_{i}, \quad q(t_{f}=1)=q_{f},
\end{equation}
where the canonical momentum is defined by $p=-\frac{3\dot{q}}{2N}$.
With the action~\eqref{eq:action} and 
the above boundary condition, the path integral 
can be exactly evaluated, and we have the following 
expression~\cite{Ailiga:2023wzl}:
\begin{align}
G[q_{f};p_{i}]&=\int_{0,-\infty}^{\infty} \mathrm{d}N \exp \biggl[\frac{iV_3}{\hbar}\biggl\{
N\left(3K+\frac{p_i^2}{3}-\Lambda q_f\right)\notag \\
&+q_f p_i-\frac{1}{3} \Lambda p_i N^2+\frac{\Lambda ^2 N^3}{9}\biggr\}\biggr]\,,
\label{eq:G-amplitude-Neumann}
\end{align}
where in the lapse integration the absence of a pole means 
there is no ambiguity of the integration contours, unlike Sec.~\ref{sec:Lefschetz-thimble-analyses}.
However, the integrand does not vanish at $N=0$ and therefore the integral along $N \in (0, \infty)$ is not guaranteed to be rewritten as a superposition of the Lefschetz thimbles, in contrast to the case of $N \in (-\infty, \infty)$.  
Thus, we consider the integration range of $N \in (-\infty, \infty)$.

We shift the lapse function $N=\bar{N}+\frac{p_i}{\Lambda}$
and this change reduces the transition amplitude to the following form:
\begin{align}
G[q_{f};p_{i}]&=\Psi(\hbar)\exp \left[\frac{iV_3}{\hbar}\frac{3}{\Lambda}\left(Kp_i+\frac{p_i^3}{27}\right)\right]\,, 
\end{align}
where 
\begin{align}
\Psi(\hbar)\equiv \int_{-\infty}^{\infty} \mathrm{d}\bar{N} \exp \left[\frac{i}{\hbar}
\left(\bar{\alpha} \bar{N}^3+ \bar{\beta} \bar{N} \right)\right]\,,
\end{align}
with $\bar{\alpha}=\frac{V_3\Lambda^2}{9}$ and 
$\bar{\beta} =V_3\left(3K-\Lambda q_f\right)$.
For instance, taking $p_i=0$, they describe the expansion of the Universe from the de Sitter throat where $q(t=0)=3/\Lambda$ to a final hypersurface with $q(t=1)= q_f$, and the 
transition amplitude takes exactly the form of the Airy function.
On the other hand, imposing the positively or negatively imaginary initial momentum $p_i= \pm 3i$, we can obtain the Vilenkin tunneling or the Hartle-Hawking no-boundary wave function, respectively, in a consistent way.

Although the Lefschetz thimble structure and resurgence analysis of 
the Airy functions are well known, by examining these analytical properties of the Airy functions, we can re-evaluate the wave function in Lorentzian quantum cosmology.
Now, by transforming $\bar{N}\to \bar{N}\hbar $, we have the following expression, 
\begin{align}
\Psi(\hbar)=\hbar\int_{-\infty}^{\infty} \mathrm{d}\bar{N}
\exp \left(i\bar{\alpha}\hbar^2 \bar{N}^3 +i\bar{\beta}\bar{N}\right)=\frac{2\pi\hbar^{\frac{1}{3}}\text{Ai}\left(\frac{\bar{\beta}}
{\sqrt[3]{3\bar{\alpha} \hbar ^2}}\right)}{\sqrt[3]{3\bar{\alpha}}} \,,
\label{eq:naive-integration-result}
\end{align} 
where we assume $\bar{\alpha}, \, \bar{\beta}>0$ in the last equation.
The integrand has two saddle points,
\begin{equation}
\bar{N}_{\pm} =\pm \frac{i}{\hbar}\sqrt{\frac{\bar{\beta}}{3\bar{\alpha}}}\,,
\end{equation}
where we define $\bar{N}_{\pm} $ taking a plus or minus sign, respectively. 
For $\bar{\alpha}, \, \bar{\beta}>0$, we have two imaginary saddle points and 
consider this case for simplicity.

Now, we consider the contributing saddle point $\bar{N}_+$
and evaluate $\Psi(\hbar)$ around this saddle point via the perturbative expansion. 
In this case, the Borel transform is given by 
\begin{equation}
{\cal B}\Psi_+(t)= 
\sum_{n=0}^\infty \frac{c_n}{\Gamma(\frac{n}{2}+1)}t^{\frac{n}{2}}\,,
\label{BorelSeries}
\end{equation}
where $c_n$ is given by 
\begin{align}
\begin{split}
c_n=&\frac{\sqrt{\hbar}}{2(3\bar{\alpha}\bar{\beta})^{\frac{1}{4}}}
 e^{-\frac{2 }{3\hbar }\sqrt{\frac{\bar{\beta}^3}{3\bar{\alpha} }}}
 \left((-1)^{n}+1\right)\\
 &\times \frac{\Gamma \left(\frac{3 n}{2}+\frac{1}{2}\right)}{\Gamma(n+1)}\left(i
\sqrt[4]{\frac{\bar{\alpha}}{27\bar{\beta}^3}}\right)^n.
\end{split}
\end{align}
The Borel transform takes the analytical form 
\begin{align}
{\cal B}\Psi_+(t)=\sqrt[4]{\frac{\pi^2\hbar^2}{3\bar{\alpha}\bar{\beta}}}e^{-\frac{2 }{3\hbar }\sqrt{\frac{\bar{\beta}^3}{3\bar{\alpha} }}} \, _2F_1\left(\frac{1}{6},\frac{5}{6};1;-\frac{3}{4}t \sqrt{\frac{3\bar{\alpha}}{\bar{\beta}^3}}\right)\,.
\end{align}
For $\bar{\alpha} >0$ and $\bar{\beta} >0$, the Borel transform has only one 
branch point along the negative real axis, and 
the Borel summation $\mathcal{S}_{\theta}\Psi_+(\hbar)$, where we take $\theta=0$, is 
\begin{align}
\begin{split}
\mathcal{S}_{\theta}\Psi_+(\hbar)&= \frac{1}{\hbar}\int_{0}^{\infty\cdot e^{i \theta}} \mathrm{d} t \, e^{-\frac{t}{\hbar}} 
{\cal B}\Psi_+(t) \\
&=\frac{2\pi\hbar^{\frac{1}{3}} \text{Ai}\left(\frac{\bar{\beta}}
{\sqrt[3]{3\bar{\alpha} \hbar ^2}}\right)}{\sqrt[3]{3\bar{\alpha}}} \,,
\end{split}
\end{align}
which is consistent with the result~\eqref{eq:naive-integration-result}. 
Thus, we show that the perturbative expansion around the contributing saddle point is 
exactly consistent with the full integration. 
In Neumann and Robin boundary conditions, the Lefschetz thimble structures of quantum cosmology are trivial, and it is possible to accurately derive the wave function of the Universe from the perturbative expansion around saddle points. 
This contrasts with the discussion of the Dirichlet conditions in Sec.~\ref{sec:Resurgence-analyses}, where a detailed analysis is necessary to see the cancellation of the ambiguities of Borel resummation.  
In both cases, the wave function of the Universe can be precisely derived 
with the resurgence theory.

\section{Conclusions}
\label{sec:conclusions}
In this paper, we studied the Lefschetz thimble structures and properties of quantum gravity corrections for the gravitational Lorentzian path integral.
We found that the thimble structures have qualitative differences among the three-parameter regions depending on the cosmological constant, extrinsic curvature, three-dimensional volume factor, and boundary conditions.

In the region (i), $12\alpha \gamma +\beta^2 >0$ with $\beta <0$, the integral is unambiguously decomposed into a superposition of the thimble integrals associated with the four real saddle points, where two of them are tunneling and the other two are no boundary ones.

In the region (ii), $12\alpha \gamma +\beta^2 <0$,  we are on the Stokes lines and the decomposition is ambiguous. 
To see what happens precisely, we studied the thimble structures for complex $\hbar$ as is usually done in the context of resurgence. 
Then, we found that the complex case does not have ambiguities of the thimble decomposition and the structures are the same between $\theta >0$ and $\theta <0$.
The $\theta =0$ case of interest can be understood as the limit $\theta \rightarrow 0^\pm$ from nonzero $\theta$. 
Therefore, we concluded that the contributing saddles in the parameter region (ii) are the two tunneling saddles around the real axis.
Our finding concludes that in traditional quantum cosmology, the tunneling wave function, rather than the no-boundary wave function, accurately describes the quantum creation of the Universe from nothing.

The case for the region (iii),  $12\alpha \gamma +\beta^2 >0$ with $\beta >0$, is also on Stokes lines.
However, we found that the contributing saddles are {\it different} between $\theta >0$ and $\theta <0$:
all four saddles in the first and third quadrants contribute for $\theta >0$, while only the two tunneling saddles in the quadrants contribute for $\theta <0$.
This implies that the Stokes multipliers for the two no-boundary saddles 
are like step functions jumping at $\theta =0$.
We found that, although each term of the summand 
may be ambiguous due to a jump at $\theta =0$, the ambiguity is canceled by that coming from the other terms, and thus the whole answer is continuous.
We explicitly demonstrated the cancellations among the ambiguities coming from the Stokes multipliers of the two no-boundary saddles and the Borel resummation of the two tunneling saddles.

It should be recalled that there is the ambiguity of the contour choice discussed in Refs.~\cite{DiazDorronsoro:2017hti,Feldbrugge:2017mbc} (see footnote~\ref{fn:contour_ambiguity}). The methodology we employed in this work can be used once we define the initial contour of $N$, but it is not what determines the contour.  Depending on the initial contour choices, the considered physical quantities can be inequivalent.  To unambiguously determine the contour to be considered, we need some additional physical inputs. We leave such considerations for future work.

For the Neumann and Robin boundary conditions, the Lefschetz thimble structures in quantum cosmology are trivial, reducing the gravitational path integral to the form of Airy functions. Consequently, deriving the no-boundary or tunneling wave function from the perturbative expansion around saddle points becomes straightforward.

This study demonstrates that applying resurgence theory to the Lorentzian quantum cosmology framework allows for a consistent derivation of the wave function of the Universe. Exploring the Lefschetz-thimble and resurgence structures in other models  
of quantum cosmology would be interesting.

\section*{Acknowledgement}
M.~H.~would like to thank Anshuman Maharana, Tomas Reis, and Ashoke Sen for useful comments on his presentation on this work at the workshop ``Non-perturbative methods in Quantum Field Theory and String Theory" at Harish-Chandra Research Institute.
H.~M.~would like to thank Shinji Mukohyama for useful discussions.
This research was conducted while M.~H.~visited the Okinawa Institute of Science and Technology (OIST) through the Theoretical Sciences Visiting Program (TSVP).
M.~H.~is supported by MEXT Q-LEAP, JST PRESTO Grant Number JPMJPR2117, Japan, JSPS Grant-in-Aid 
for Transformative Research Areas (A) ``Extreme Universe" JP21H05190 [D01].
This work is supported by Japan Society for the Promotion of Science (JSPS) KAKENHI Grant Numbers JP22H01222 (M.~H.), JP22KJ1782 (H.~M.), JP23K13100 (H.~M.), and JP23KJ1162 (K.~O.).
This work was supported by IBS under the project code, IBS-R018-D1.

\bibliographystyle{utphys}
\bibliography{Refs.bib}

\end{document}